\documentclass[]{imag-ms-template}
\usepackage{multicol,multirow}
\usepackage{subcaption}
\usepackage{caption}
\usepackage{xcolor}
\usepackage{diagbox}
\usepackage{array}
\usepackage{booktabs}
\usepackage{hhline}
\usepackage{colortbl}

\usepackage{rotating} % rotate the table 90 degrees
\usepackage{pdflscape}
\usepackage{tabularray} 
\usepackage{array}

\title{Random Effects Models for Understanding Variability and Association between Brain Functional and Structural Connectivity}

\author{Lingyi Peng,$^{1\dagger}$ Qiaochu Wang,$^{2\dagger}$ Yaotian Wang,$^{3}$ Jie He,$^{2}$ Xu Zou,$^{2}$\\ Shuoran Li,$^{2}$ Dana L. Tudorascu,$^{1}$ David J. Schaeffer,$^{4}$ Lauren Schaeffer,$^{4}$ \\Diego Szczupak,$^{4}$ Emily S. Rothwell,$^{4}$ Stacey J. Sukoff Rizzo,$^{4}$ \\Gregory W. Carter,$^{5}$ Afonso C. Silva,$^{4}$ and Tingting Zhang$^{2\ast}$\\
{\small $^{1}$Department of Biostatistics and Health Data Science, University of Pittsburgh,}\\
{\small 130 De Soto Street, Pittsburgh, PA 15261, USA}\\
{\small $^{2}$Department of Statistics, University of Pittsburgh,}\\
{\small 230 S Bouquet Street, Pittsburgh, PA 15260, USA}\\
{\small $^{3}$Department of Biostatistics and Bioinformatics, Emory University,}\\
{\small 1518 Clifton Rd. NE, Atlanta, GA 30322, USA}\\
{\small $^{4}$Department of Neurobiology, University of Pittsburgh, }\\
{\small 200 Lothrop Street, Pittsburgh, PA 15213, USA}\\
{\small $^{5}$The Jackson Laboratory, 600 Main Street, Bar Harbor, ME 04609, USA}\\
{\small $^\dagger$Lingyi Peng and Qiaochu Wang are equally contributing authors}\\
{\small $^\ast$Corresponding author: Tingting Zhang, 1826 Wesley W. Posvar Hall, Room 108, }\\
{\small University of Pittsburgh, Pittsburgh, PA 15260, USA. Email: tiz67@pitt.edu}\\
}

\begin{document} 

\maketitle


\begin{abstract}
The human brain is organized as a complex network, where connections between regions are characterized by both functional connectivity (FC) and structural connectivity (SC). While previous studies have primarily focused on network-level FC-SC correlations (i.e., the correlation between FC and SC across all edges within a predefined network), edge-level correlations (i.e., the correlation between FC and SC across subjects at each edge) has received comparatively little attention.  In this study, we systematically analyze both network-level and edge-level FC-SC correlations, demonstrating that they lead to divergent conclusions about the strength of brain function-structure association. To explain these discrepancies, we introduce new random effects models that decompose FC and SC variability into different sources: subject effects, edge effects, and their interactions. Our results reveal that network-level and edge-level FC-SC correlations are influenced by different effects, each contributing differently to the total variability in FC and SC. This modeling framework provides the first statistical approach for disentangling and quantitatively assessing different sources of FC and SC variability and yields new insights into the relationship between  functional and structural brain networks.
\end{abstract}

\section{Introduction}

The human brain functions as a complex, high-dimensional network in which nodes represent distinct brain regions and edges correspond to connections between them. Each connection between a pair of regions is characterized by both functional connectivity (FC) and structural connectivity (SC). FC refers to the statistical or temporal association between activity of a pair of brain regions \citep{friston1994functional,friston2011functional}, whereas SC denotes the white matter fiber tracts that physically link the pair of regions \citep{Hagmann2008Mapping,RubSporns2010,Sporns2011}. Since SC serves as the brain's physical architecture and provides the substrate through which neural signals propagate, understanding how this structural framework (SC) supports FC is a fundamental question in neuroscience \citep{Sporns04,sporns2005human,Sporns2011,hermundstad2013structural,mivsic2016network,van2013network}.

To address this question, researchers have developed a range of modeling and predictive approaches aimed at uncovering the relationship between SC and FC. Specifically, biophysical models, especially neural mass models, have been constructed to explain the brain's neuronal activity at the macroscopic scale and provide insights into how brain functional patterns may emerge from SC \citep{deco2013resting,hansen2015functional,wang2019inversion, honey2009predicting, breakspear2017dynamic}. Additionally, numerous prediction methods have been introduced to estimate FC from SC. These include methods based on spectral graph theory \citep{abdelnour2018functional, abdelnour2014network, atasoy2016harmonic, becker2018spectral,benkarim2022riemannian}, network communication approaches \citep{avena2018communication, goni2014resting, Misic2015Cooperative, zamani2022local, seguin2020network}, and deep learning methods \citep{chen2024group, sarwar2021structure, li2019mapping, neudorf2022structure,yang2022interpretable}. While most studies focus on predicting FC from SC, there is also a growing interest in predicting SC from FC \citep{zhang2022predicting, wang2020understanding}.

In addition to biophysical modeling and predictive approaches, a widely adopted method for investigating the relationship between FC and SC involves directly assessing their correlations. Given the complex network organization of the brain, there exist two distinct types of FC-SC correlations, each offering a unique perspective of the relationship between FC and SC. 

The first type is the network-level FC-SC correlation  (hereafter, network correlation),
which quantifies the association between FC and SC across all edges or a selected subset of edges within a specific brain network. This network can belong to an individual subject or be the population-mean network. This network correlation offers insights into the overall function-structure coupling in a specific network.

\begin{figure}[h!]
\begin{center}
    \includegraphics[width=\textwidth]{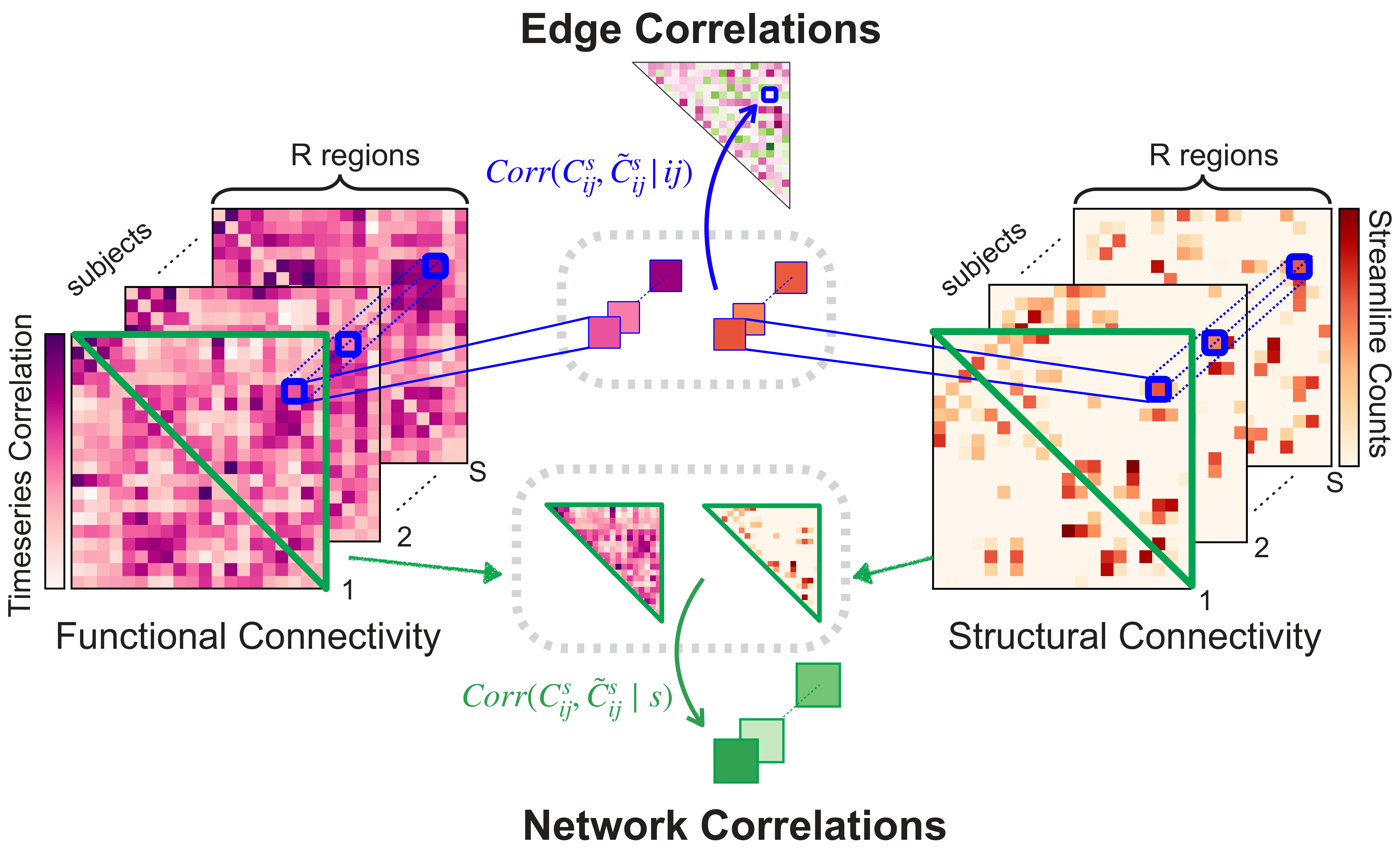}
    \caption{Workflow for calculating network and edge correlations. Each of $S$ subjects is associated with an $R$-by-$R$ symmetric FC matrix (purple, left) and an $R$-by-$R$ symmetric SC matrix (orange, right), where each matrix entry represents the connectivity measure between a pair of regions (i.e., of a network edge). For each subject, the network FC-SC correlation is calculated as the Pearson correlation between the upper-triangular  (excluding the diagonal) entries of their FC and SC matrices, as illustrated in the lower part of the figure. For each network edge, the edge correlation is computed by correlating the FC and SC values of the edge across $S$ subjects, as illustrated in the upper part.}
\label{fig: SC-FC Correlations}
\end{center}
\end{figure}

The second is the edge-level FC-SC correlation (edge correlation), which is the correlation between FC and SC at a given network edge (i.e., between a specific pair of regions) across a group of subjects. This correlation measures how strongly SC and FC co-vary at that particular connection throughout the population. As illustrated in Figure \ref{fig: SC-FC Correlations}, subjects' network correlations capture individual differences in overall network function-structure coupling, while edge correlations reveal how FC-SC associations of a population vary across edges (pairs of brain regions).

Most prior studies have focused on subjects' network correlations \citep{greicius2009resting,toosy2004characterizing,huang2016linking}, consistently reporting moderate positive correlations between individual subjects’ functional and structural networks, as well as between population-mean functional and structural networks \citep{honey2009predicting,straathof2019systematic,mivsic2016network,skudlarski2008measuring}. However, substantial regional variability in these correlations has been observed across different functional brain networks \citep{honey2010can}. For example, the default mode network (DMN), visual network, and sensorimotor network (SMN) tend to show stronger network FC-SC correlations than other networks \citep{baum2020development,honey2009predicting,gu2021heritability,zamani2022local,skudlarski2008measuring,Vázquez2019gradients}. This heterogeneity prompts a more thorough examination of FC-SC correlations across various regions, particularly edge correlations.

Although there is an increasing interest in understanding FC-SC relationships, studies focusing on edge correlations are less prevalent than those concerning network correlations \citep{faskowitz2022edges,faskowitz2021edge}. Available studies suggest that these correlations are generally weak \citep{reddi2017understanding,mollink2019spatial}, indicating that different mechanisms or sources of variability influence these two types of correlations. Yet, there has been insufficient effort to systematically explore and identify these underlying causes, thus maintaining a significant gap in our understanding of the relationship between functional and structural brain networks.

In this work, we address the gap through a comprehensive comparison of network and edge FC-SC correlations.  Our results confirm that these two types of correlations lead to divergent conclusions regarding the strength of FC-SC associations. To explain these discrepancies, we introduce a new random effects modeling framework that partitions FC and SC variability into several components: (1) subject effects, capturing variability across subjects; (2) edge effects, reflecting variability across edges; (3) interaction effects, describing variability across edges within a single subject or across subjects for a specific edge; and (4) residuals, accounting for unexplained variability. Using this framework, we perform variance decomposition to quantify the relative contribution of each component to the overall variability in FC and SC.

By disentangling the distinct sources of variability in FC and SC, our approach clarifies why network and edge FC-SC correlations yield differing results---namely, because they reflect different underlying components of FC and SC variability. Furthermore, our modeling approach quantifies the proportion of variability  attributable to each effect, enabling direct comparisons of their relative influences. Overall, this modeling-based approach provides a first statistical framework to investigate and quantify different sources of variability observed in the brain networks measured by different imaging modalities and offers new insights into the relationship between functional and structural brain networks.

\section{Materials and Methods} \label{sec:methods}

\subsection{Data Acquisition and Preprocessing}
We analyzed resting-state functional MRI (rs-fMRI) and diffusion MRI (dMRI) data of 981 healthy adults (age 22–37 years; 458 males, 523 females) from the Human Connectome Project (HCP) \citep{van2013wu}. All participants gave written informed consent in accordance with protocols approved by the Institutional Review Board of Washington University in St. Louis. 
To ensure data completeness and consistency, we included only subjects who completed all four rs-fMRI runs and the full dMRI protocol.

All imaging data were acquired on a customized 3T Siemens Connectome-Skyra scanner using multiband sequences optimized for high-resolution imaging. The rs-fMRI protocol included four 15-minute runs (1200 volumes per run; TR = 720 ms; 2 mm isotropic resolution), with two runs collected in each of two sessions using alternating left-right and right-left phase-encoding directions \citep{smith2013resting}. The dMRI scanning protocol used 270 diffusion-weighted directions distributed across three b-value shells (b = 1000, 2000, and 3000 s/mm²), along with multiple b = 0 images acquired at 1.25 mm isotropic resolution \citep{sotiropoulos2013advances}.

We used the HCP minimally preprocessed rs-fMRI data \citep{glasser2013minimal}, which had undergone structural alignment, motion correction, surface-based mapping, and denoising using tools from FSL \citep{jenkinson2012fsl}, FreeSurfer \citep{fischl2012freesurfer}, Connectome Workbench \citep{marcus2011informatics}, and ICA-FIX \citep{smith2013resting,salimi2014automatic, griffanti2014ica}.

For dMRI, we also used HCP minimally preprocessed data, which had been corrected for diffusion-specific artifacts, including eddy currents, head motion, and EPI distortions, and registered to each subject’s native T1-weighted anatomical space \citep{glasser2013minimal}. Following this, we applied an additional bias field correction using the N4ITK algorithm \citep{tustison2010n4itk}.

\subsection{Functional Network Construction}

The cortex was divided into 200 distinct regions using the Schaefer-200 parcellation \citep{schaefer2018local}. For each region, we averaged the fMRI signals of voxels within the region to derive a single representative time series. Then, for each subject, we computed FC between each region pair as the Pearson correlation coefficient between the fMRI time series of the region pair. Subsequently, we applied a Fisher z-transformation to the resulting coefficients to stabilize variance and improve normality. This procedure produced a symmetric 200-by-200 FC matrix per subject, with each entry corresponding to the Fisher z-transformed correlation between fMRI data of two brain regions.

\subsection{Structural Network Construction}

We constructed structural  networks using preprocessed diffusion MRI data and the MRtrix3 software package \citep{tournier2019mrtrix3}. Brain regions were also defined by the Schaefer-200 parcellation \citep{schaefer2018local}.  For each subject, we first estimated fiber orientation distributions (FODs) using multi-shell, multi-tissue constrained spherical deconvolution \citep{jeurissen2014multi}. Based on these estimated FODs, whole-brain deterministic (SD\_STREAM, \cite{tournier2012mrtrix}), anatomically constrained tractography (ACT) was performed. The tractography algorithm was conducted with dynamic seeding, generation of 5 million streamlines, an FOD amplitude cutoff of 0.05, and a maximum streamline length of 300 mm. We computed streamline weights using the Spherical-deconvolution Informed Filtering of Tractograms (SIFT2) algorithm \citep{smith2015sift2}.

For each subject, SC was quantified by summing the SIFT2-derived streamline weights connecting each pair of brain regions. To reduce the  right skewness in the distribution of streamline counts and improve comparability across connections and subjects, we applied a logarithmic transformation to the resulting values \citep{luppi2021combining}. This yielded a 200-by-200 symmetric SC matrix for each subject, where each entry represents the log-transformed, SIFT2-weighted streamline count between two brain regions.

In summary, the default setting used for constructing functional and structural networks includes (1) the Schaefer-200 parcellation to define brain regions, (2) deterministic tractography to generate streamlines, and (3) SC estimation via log-transformed sums of SIFT2-weighted streamlines. The resulting FC and SC matrices serve as the basis for all subsequent FC-SC correlation analyses.

\subsection{Mathematical Definitions of FC-SC Correlations}

We present the definitions and mathematical formulations for individual subjects' network FC-SC correlations, group-average network FC-SC correlations, and edge FC-SC correlations below.

\textbf{Subject network FC-SC correlation.} Let $C_{ij}^s$ and $\tilde{C}_{ij}^s$ denote the FC and SC values, respectively, of edge $(i,j)$ connecting brain regions $i$ and $j$ for subject $s$, where $i,j=1,\ldots,R$ and $s=1,\ldots,S$. Here, $R$ represents the total number of brain regions, and $S$ is the number of subjects under analysis. The network correlation of subject $s$ was computed as the Pearson correlation between FC and SC values across all edges in the subject’s brain network:
\begin{equation}\label{eq:NetworkCorr}
Corr(C_{ij}^s,\tilde{C}_{ij}^s \mid s)=\frac{\sum_{(i, j) \in \Omega}(C_{ij}^s-C_m^s)(\tilde{C}_{ij}^s-\tilde{C}_m^s)}{\sqrt{\sum_{(i,j) \in \Omega}(C_{ij}^s-C_m^s)^2}\sqrt{\sum_{(i,j) \in \Omega}(\tilde{C}_{ij}^s-\tilde{C}_m^s)^2}}, \end{equation} where $\Omega$ denotes the set of all edges in the network ($|\Omega|=19{,}900$ for the Schaefer-200 parcellation). In addition, $C_m^s$ and $\tilde{C}_m^s$ denote the mean FC and SC across all network edges of subject $s$, respectively: $$C_m^s=\frac{\sum_{(i, j) \in \Omega} C_{ij}^s}{|\Omega|}~~\mbox{and}~~\tilde{C}_m^s=\frac{\sum_{(i, j) \in \Omega} \tilde{C}_{ij}^s}{|\Omega|}.$$

\textbf{Group-average network FC-SC correlation.} We first compute group-average functional and structural networks. Specifically, for edge $(i,j)$, the group-average FC and SC, respectively, are:
$${C}_{ij}^m=\frac{\sum_{s\in S}{C}_{ij}^s}{S}~~\mbox{and}~~  \tilde{C}_{ij}^m=\frac{\sum_{s\in S}{\tilde{C}}_{ij}^s}{S}.$$
The group-average network FC-SC correlation is: \begin{equation}\label{eq:AveNetworkCorr}Corr(C^m_{ij},\tilde{C}^m_{ij})=\frac{\sum_{(i, j) \in \Omega}(C^m_{ij}-\langle C\rangle)(\tilde{C}^m_{ij}-\langle \tilde{C}\rangle)}{\sqrt{\sum_{(i,j) \in \Omega}(C^m_{ij}-\langle C\rangle)^2}\sqrt{\sum_{(i,j) \in \Omega}(\tilde{C}^m_{ij}-\langle \tilde{C}\rangle)^2}}, \end{equation}
where $\langle C\rangle=\sum_{(i, j) \in \Omega}{C}_{ij}^m/|\Omega|$ and $\langle \tilde{C}\rangle=\sum_{(i, j) \in \Omega}{\tilde{C}}_{ij}^m/|\Omega|$.

\textbf{Edge FC-SC correlation.}  The FC-SC correlation of edge $(i,j)$ is computed as:\begin{equation}\label{eq:EdgeCorr}Corr(C_{ij}^s,\tilde{C}_{ij}^s \mid ij)=\frac{\sum_{s=1}^S (C_{ij}^s-C_{ij}^m)(\tilde{C}_{ij}^s-\tilde{C}_{ij}^m)}{\sqrt{\sum_{s=1}^S(C_{ij}^s-C_{ij}^m)^2}\sqrt{\sum_{s=1}^S(\tilde{C}_{ij}^s-\tilde{C}_{ij}^m)^2}}. \end{equation} 
\subsection{Significance of FC-SC Correlations}
We assessed the statistical significance of FC-SC correlations using Pearson correlation tests. We reported the p-value for the group-average network correlation. For edge correlations across all edges and subject network correlations across all subjects, we applied false discovery rate (FDR) correction and reported FDR-adjusted p-values \citep[q-values]{benjamini1995controlling}. A significance threshold of 0.05 was consistently used for both p-values and q-values.

\subsection{Random Effects Models for FC and SC}
We proposed random effects models for FC and SC of different network edges and subjects:
\begin{eqnarray}
C_{ij}^s&=&\mu+{\alpha}_{ij}+\beta^s+\eta_{ij}\cdot\varpi^s+\epsilon^s_{ij},\label{eq:FC}\\
\tilde{C}_{ij}^s&=&\tilde{\mu}+\tilde{\alpha}_{ij}+\tilde{\beta}^{s}+\tilde{\eta}_{ij}\cdot \tilde{\varpi}^s+ \tilde{\epsilon}^s_{ij}  ,\label{eq:SC} \end{eqnarray}
where $\mu$ and $\tilde{\mu}$ are constants representing the global means of FC and SC, respectively, across all subjects and edges. The other components are random variables that capture different sources of variability:
\begin{itemize}
\item Main edge effects: $\alpha_{ij}$ and $\tilde{\alpha}_{ij}$, respectively, represent the population-mean FC and SC deviations from the global mean specific to edge $(i,j)$. That is, $\mu + \alpha_{ij}$ and $\tilde{\mu} + \tilde{\alpha}_{ij}$ correspond to the population-mean FC and SC values for edge $(i,j)$. These effects capture population-mean FC and SC variability across network edges.
\item Main subject effects: $\beta^s$ and $\tilde{\beta}^s$, respectively, denote subject-specific FC and SC deviations from the global mean, averaged across all edges. Hence, $\mu + \beta^s$ and $\tilde{\mu} + \tilde{\beta}^s$ represent network-average FC and SC values specific to subject $s$. These effects quantify variability in mean network connectivity across subjects.
\item Interaction effects: $\eta_{ij}\cdot\varpi^s$ and $\tilde{\eta}_{ij}\cdot \tilde{\varpi}^s$, respectively, capture how FC and SC variability differ across edges within each subject or across subjects for a specific edge. We refer to $\eta_{ij}$ and $\tilde{\eta}_{ij}$ as interaction edge effects in FC and SC, respectively, and refer to $\varpi^s$ and $\tilde{\varpi}^s$ as interaction subject effects in FC and SC, respectively. 
\item The residual terms: $\epsilon^s_{ij}$ and $\tilde{\epsilon}^s_{ij}$, respectively, represent the remaining variability in FC and SC not accounted by the main or interaction effects.
\end{itemize}

To enable straightforward interpretation and ensure that each effect’s contribution can be independently assessed, all random components within each model, as well as the components corresponding to different effects across models, are assumed to be mutually independent:
 \begin{equation}\label{eq:Indp}
 \alpha_{ij}\perp \beta^s\perp \eta_{ij}\perp \varpi^s\perp\epsilon^s_{ij} ~~~\mbox{and}~~ \tilde{\alpha}_{ij}\perp \tilde{\beta}^s\perp \tilde{\eta}_{ij}\perp\tilde{\varpi}^s\perp\tilde{\epsilon}^s_{ij}~~~\mbox{and}\end{equation}
 \begin{equation}\label{eq:Indp2}
(\alpha_{ij},\tilde{\alpha}_{ij})\perp (\beta^s ,\tilde{\beta}^s)\perp(\eta_{ij},\tilde{\eta}_{ij})\perp (\varpi^s,\tilde{\varpi}^s)\perp \epsilon^s_{ij}\perp\tilde{\epsilon}^s_{ij} .\end{equation}
These independence assumptions imply that only the corresponding effects in FC and SC (e.g., $\alpha_{ij}$ and $\tilde{\alpha}_{ij}$ for main edge effects) may be correlated with each other, while remaining independent of other effects of variability. As a result, FC-SC associations under the random effects models \eqref{eq:FC} and \eqref{eq:SC} depend on four key correlations between corresponding effects in FC and SC: 
\begin{equation}\label{eq:CorrEffect}
\rho_{\boldsymbol{\alpha}}=Corr(\alpha_{ij}, \tilde{\alpha}_{ij}),~~\rho_{\boldsymbol{\beta}}=Corr(\beta^s, \tilde{\beta}^s), ~~\varrho_{\boldsymbol{\eta}}=Corr(\eta_{ij}, \tilde{\eta}_{ij}),~~\mbox{and}~~\varrho_{\boldsymbol{\varpi}}=Corr(\varpi^s, \tilde{\varpi}^s),\end{equation}
where $\boldsymbol{\alpha}=\{\alpha_{ij}, \tilde{\alpha}_{ij},i< j\in \{1,\ldots, R\}\}$, $\boldsymbol{\beta}=\{\beta^s, \tilde{\beta}^s,s=1,\ldots, S\}$, $\boldsymbol{\eta}=\{\eta_{ij}, \tilde{\eta}_{ij},i< j\in \{1,\ldots, R\}\}$, and $\boldsymbol{\varpi}=\{\varpi^s, \tilde{\varpi}^s,s=1,\ldots, S\}$. 

Finally, each random component in the models \eqref{eq:FC} and \eqref{eq:SC} is assumed to be independently and identically distributed (i.i.d.) with zero mean and specific variances:\begin{eqnarray}
\alpha_{ij}&\stackrel{i.i.d}{\sim} &\mathcal{D}(0,\tau_{\boldsymbol{\alpha}}^2),
~~\beta^s\stackrel{i.i.d}{\sim} \mathcal{D}(0,\tau_{\boldsymbol{\beta}}^2),
~~\eta_{ij}\stackrel{i.i.d}{\sim}\mathcal{D}(0,\sigma^2_{\boldsymbol{\eta}}),~~ \varpi^s \stackrel{i.i.d}{\sim}\mathcal{D}(0,\sigma_{\boldsymbol{\varpi}}^2),~~ \epsilon^s_{ij}\stackrel{i.i.d}{\sim} \mathcal{D}(0,\sigma_{\boldsymbol{\epsilon}}^2); \label{eq:FCVar}\\
\tilde{\alpha}_{ij}&\stackrel{i.i.d}{\sim}& \mathcal{D}(0,\tilde{\tau}_{\boldsymbol{\alpha}}^2),~~\tilde{\beta}^s\stackrel{i.i.d}{\sim} \mathcal{D}(0,\tilde{\tau}_{\boldsymbol{\beta}}^2),
~~\tilde{\eta}_{ij}\stackrel{i.i.d}{\sim}\mathcal{D}(0,\tilde{\sigma}^2_{\boldsymbol{\eta}}),~~ \tilde{\varpi}^s \stackrel{i.i.d}{\sim}\mathcal{D}(0,\tilde{\sigma}_{\boldsymbol{\varpi}}^2),~~ \tilde{\epsilon}^s_{ij}\stackrel{i.i.d}{\sim} \mathcal{D}(0,\tilde{\sigma}_{\boldsymbol{\epsilon}}^2),\label{eq:SCVar}
\end{eqnarray}
for $i,j=1,\ldots, R$, $s=1,\ldots,S$, where $\mathcal{D}(\mu,\sigma^2)$ denotes a distribution function with mean $\mu$ and variances $\sigma^2$. To avoid the identifiability issue, we fixed the variances of $\eta_{ij}$ and  $\tilde{\eta}_{ij}$ to be 1: $\sigma^2_{\boldsymbol{\eta}}=\tilde{\sigma}^2_{\boldsymbol{\eta}}=1.$

\subsection{Model Parameter Estimates}\label{sec:ParaEst}

In the following, the notation $\hat{\theta}$ represents the estimate of a parameter
$\theta$ related to FC and $\breve{\theta}$ denotes the corresponding estimate of the parameter $\tilde{\theta}$ for SC.

The estimates of parameters $\mu$, $\alpha_{ij}$, $\beta^s$, $\eta_{ij}$, and $\varpi^s$ in the model \eqref{eq:FC} are obtained through minimizing the sum of squared errors:
$$\sum_{i< j}^R\sum_{s=1}^S(C^s_{ij}-\mu-\alpha_{ij}-\beta^s-\eta_{ij}\cdot \varpi^s)^2$$ subject to the following identifiability constraints:
$$\sum_{i< j}^R\alpha_{ij}=\sum_{s=1}^S\beta^s=\sum_{i< j}^R\eta_{ij}=\sum_{s=1}^S \varpi^s=0~~\mbox{and}~~ 
\sum_{i< j}^R\frac{\eta^2_{ij}}{R\cdot (R-1)/2}=1.$$
Closed-form estimates under these constraints are:
$$\hat{\mu}= \sum_{i< j}^R\sum_{s=1}^S\frac{C^s_{ij}}{S\cdot R\cdot(R-1)/2},~~\hat{\alpha}_{ij}=\sum_{s=1}^S\frac{C^s_{ij}}{S}-\hat{\mu},~~\mbox{and}~~\hat{\beta}^s=\sum_{i< j}^R\frac{C^s_{ij}}{R\cdot(R-1)/2}-\hat{\mu}.$$ 
The remaining parameters $\eta_{ij}$ and $\varpi^s$ are estimated numerically using the Newton-Raphson optimization algorithm. The ensuing estimates are denoted by $\hat{\eta}_{ij}$ and $\hat{\varpi}^s$. The residuals in the FC model \eqref{eq:FC} are then computed as:$$\hat{\epsilon}^s_{ij}=C_{ij}^s-\hat{\mu}-\hat{\alpha}_{ij}-\hat{\beta}^s-\hat{\eta}_{ij}\cdot \hat{\varpi}^s.$$ 

Based on these parameter estimates, we subsequently estimate the variances of the random components in \eqref{eq:FCVar} using the variances of their respective parameter estimates. For example, $\tau^2_{\boldsymbol{\alpha}}$ is estimated by the variance of $\{\hat{\alpha}_{ij} , i< j\}$. The estimates of variances in \eqref{eq:FCVar} are denoted as $\hat{\tau}^2_{\boldsymbol{\alpha}}$, $\hat{\tau}^2_{\boldsymbol{\beta}}$, $\hat{\sigma}^2_{\boldsymbol{\eta}}$, $\hat{\sigma}^2_{\boldsymbol{\varpi}}$, and
$\hat{\sigma}^2_{\boldsymbol{\epsilon}}$.

The same procedure is applied to the SC model \eqref{eq:SC}. The estimates of $\tilde{\mu}$, $\tilde{\alpha}_{ij}$, and $\tilde{\beta}^s$ are
$$\breve{\mu}= \sum_{i< j}^R\sum_{s=1}^S\frac{\tilde{C}^s_{ij}}{S\cdot R\cdot(R-1)/2},~~\breve{\alpha}_{ij}=\sum_{s=1}^S\frac{\tilde{C}^s_{ij}}{S}-\breve{\mu},~~\mbox{and}~~\breve{\beta}^s=\sum_{i< j}^R\frac{\tilde{C}^s_{ij}}{R\cdot(R-1)/2}-\breve{\mu}.$$

As with FC, the SC interaction terms $\tilde{\eta}_{ij}$ and $\tilde{\varpi}^s$ are  estimated numerically,  resulting in estimates $\breve{\eta}_{ij}$ and $\breve{\varpi}^s$. The SC residuals are then computed as:$$\breve{\epsilon}^s_{ij}=\tilde{C}_{ij}^s-\breve{\mu}-\breve{\alpha}_{ij}-\breve{\beta}^s-\breve{\eta}_{ij}\cdot \breve{\varpi}^s.$$ 
The variances of the random components in \eqref{eq:SCVar} are estimated by the variances of the corresponding parameter estimates, which are denoted by $\breve{\tau}^2_{\boldsymbol{\alpha}}$, $\breve{\tau}^2_{\boldsymbol{\beta}}$, $\breve{\sigma}^2_{\boldsymbol{\eta}}$, $\breve{\sigma}^2_{\boldsymbol{\varpi}}$, and
$\breve{\sigma}^2_{\boldsymbol{\epsilon}}$.

Finally, we assess the correlations between FC and SC effects by computing correlations between their respective parameter estimates:
$$\hat{\rho}_{\boldsymbol{\alpha}},~~ \hat{\rho}_{\boldsymbol{\beta}},~~ \hat{\varrho}_{\boldsymbol{\eta}}, ~\mbox{and}~ \hat{\varrho}_{\boldsymbol{\varpi}},$$
which represent empirical estimates of the correlations, 
$\rho_{\boldsymbol{\alpha}}$, $\rho_{\boldsymbol{\beta}}$, $\varrho_{\boldsymbol{\eta}}$, and $\varrho_{\boldsymbol{\varpi}}$, defined in Equation \eqref{eq:CorrEffect}.

\subsection{Variance Decomposition of Total FC and SC Variability}\label{Sec:VarDec} We decompose the total variability of FC and SC in the data into distinct components corresponding to the main edge effects, main subject effects, interaction effects, and residual terms, as specified in the models \eqref{eq:FC} and \eqref{eq:SC}. This variance decomposition allows us to quantify the relative contribution of each source of variability to the overall observed variability in FC and SC across subjects and edges. 

Specifically, the total variability of FC in the data is measured by the total sum of squares across all edges and subjects, denoted as $SST$:
$$SST=\sum_{i< j}^R\sum_{s=1}^S(C^s_{ij}-\hat\mu)^2$$ 

This total sum of squares is decomposed into four components corresponding to the main edge effects, main subject effects, interaction effects, and residual terms in the model \eqref{eq:FC}:
\begin{align*}
SST%&=\sum_{i\leq j}^R\sum_{s=1}^S(C^s_{ij}-\hat\mu)^2\\
%=\sum_{i\leq j}^R\sum_{s=1}^S(\hat{\mu}+\hat{\alpha}_{ij}+\hat{\beta}^s+\hat{\eta}_{ij}\cdot\hat{\varpi}^s+\hat{\epsilon}_{ij}^s-\hat\mu)^2\\
%          &=\sum_{i\leq j}^R\sum_{s=1}^S(\hat{\alpha}_{ij}+\hat{\beta}^s+\hat{\eta}_{ij}\cdot\hat{\varpi}^s+\hat{\epsilon}_{ij}^s)^2\\
%          &=\sum_{i\leq j}^R\sum_{s=1}^S(\hat{\alpha}_{ij})^2+\sum_{i\leq j}^R\sum_{s=1}^S(\hat{\beta}^s)^2+\sum_{i\leq j}^R\sum_{s=1}^S(\hat{\eta}_{ij}\cdot\hat{\varpi}^s)^2+\sum_{i\leq j}^R\sum_{s=1}^S(\hat{\epsilon}_{ij}^s)^2\\
          =SS({\boldsymbol{\alpha}})+SS({{\boldsymbol{\beta}}})+SS({\boldsymbol{\eta\cdot{\varpi}}})+SS({\boldsymbol{\epsilon}})
\end{align*}
where each sum of squares is defined as: 
\begin{itemize}
\item FC variability attributed to the main edge effects:
$SS({\boldsymbol{\alpha}}) = S \sum_{i <j}^R (\hat{\alpha}_{ij})^2$,
\item FC variability attributed to the main subject effects:
$ SS({{\boldsymbol{\beta}}}) = \frac{R\cdot (R-1)}{2} \sum_{s=1}^S (\hat{\beta}^s)^2$,
\item FC variability attributed to the interaction effects:
$SS({\boldsymbol{\eta\cdot{\varpi}}}) = \sum_{i < j}^R \sum_{s=1}^S (\hat{\eta}_{ij}\cdot\ \hat{\varpi}^s)^2$, and
\item FC variability attributed to the residual terms:
 $SS({\boldsymbol{\epsilon}}) = \sum_{i < j}^R\sum_{s=1}^S (\hat{\epsilon}^s_{ij})^2.$
\end{itemize}
Given the above decomposition, the proportions of $SST$ explained by the four components are:
$$R^2(\boldsymbol{\alpha})=\frac{SS({\boldsymbol{\alpha}})}{SST},~~R^2({\boldsymbol{\beta}})=\frac{SS({\boldsymbol{\beta}})}{SST},~~R^2({\boldsymbol{\eta\cdot{\varpi}}})=\frac{SS({\boldsymbol{\eta\cdot{\varpi}}})}{SST},~~\mbox{and}~~R^2({\boldsymbol{\epsilon}})=\frac{SS({\boldsymbol{\epsilon}})}{SST}. $$

We perform the same variance decomposition of the total sum of squares for SC, denoted as $\tilde{SST}$:
$$\tilde{SST}=\sum_{i< j}^R\sum_{s=1}^S(\tilde{C}^s_{ij}-\breve\mu)^2=\tilde{SS}(\boldsymbol{\alpha})+\tilde{SS}({\boldsymbol{\beta}})+\tilde{SS}({\boldsymbol{\eta}\cdot\boldsymbol{\varpi}})+\tilde{SS}({\boldsymbol{\epsilon}}),$$
where the four components are defined as: 
\begin{itemize}
\item SC variability attributed to the main edge effects:
$\tilde{SS}({\boldsymbol{\alpha}}) =  \sum_{i < j}^R (\breve{\alpha}_{ij})^2$,
\item SC variability attributed to the main subject effects:
$ \tilde{SS}({{\boldsymbol{\beta}}}) = \frac{R\cdot(R-1)}{2} \sum_{s=1}^S (\breve{\beta}^s)^2$,
\item SC variability attributed to the interaction effects:
$\tilde{SS}({\boldsymbol{\eta\cdot{\varpi}}}) = \sum_{i < j}^R \sum_{s=1}^S (\breve{\eta}_{ij}\cdot\ \breve{\varpi}^s)^2$, and
\item SC variability attributed to the residual terms:
 $\tilde{SS}({\boldsymbol{\epsilon}}) = \sum_{i < j}^R\sum_{s=1}^S (\breve{\epsilon}^s_{ij})^2.$
\end{itemize}
The proportions of the total SC variability explained by each component in the model \eqref{eq:SC} are given by:
$$
\tilde{R}^2(\boldsymbol{\alpha})=\frac{\tilde{SS}({\boldsymbol{\alpha}})}{\tilde{SST}},~~\tilde{R}^2({\boldsymbol{\beta}})=\frac{\tilde{SS}({{\boldsymbol{\beta}}})}{\tilde{SST}},~~\tilde{R}^2({\boldsymbol{\eta}\cdot{\boldsymbol{\varpi}}})=\frac{\tilde{SS}({\boldsymbol{\eta}\cdot{\boldsymbol{\varpi}}})}{\tilde{SST}},~~\mbox{and}~~\tilde{R}^2({\boldsymbol{\epsilon}})=\frac{\tilde{SS}({\boldsymbol{\epsilon}})}{\tilde{SST}}. 
$$

\section{Results}\label{sec:results}

\subsection{Subject Network FC-SC Correlations}

The subject network FC-SC correlations, $Corr(C_{ij}^s,\tilde{C}_{ij}^s\mid s)$ in \eqref{eq:NetworkCorr}, are moderate positive for all subjects, ranging from 0.12 to 0.31. The histogram of these correlations is presented in Figure  \ref{fig:default_setting_correlations}.\textit{A}. The average of these correlations is 0.23, with a standard deviation (SD) of 0.032 (Table \ref{tab:fcsc_correlation_summary}.\textit{(A)}).  All of these correlations are statistically significant at the 5\% FDR threshold. 

\subsection{Group-Average Network FC-SC Correlation} 
The group-average network FC-SC correlation, as defined in \eqref{eq:AveNetworkCorr}, is 0.331 and statistically significant (bootstrap p-value $<$ 0.001). This value is higher than all the subject network correlations, as illustrated in Figure \ref{fig:default_setting_correlations}.\textit{A}. Previous studies \citep{honey2009predicting,straathof2019systematic} also reported this pattern of higher group-average correlation compared to subject network correlations.

\begin{table}
\centering
\setlength{\extrarowheight}{0pt}
\addtolength{\extrarowheight}{\aboverulesep}
\addtolength{\extrarowheight}{\belowrulesep}
\setlength{\aboverulesep}{0pt}
\setlength{\belowrulesep}{0pt}
\label{tab:fcsc_correlation_summary}
\arrayrulecolor{black}
\begin{tabular}{>{\hspace{0pt}}m{0.3\linewidth}!{\color{black}\vrule}>{\centering\hspace{0pt}}m{0.18\linewidth}>{\centering\hspace{0pt}}m{0.18\linewidth}!{\color{black}\vrule}>{\centering\arraybackslash\hspace{0pt}}m{0.18\linewidth}} 
\toprule
\textbf{Setting} & \textbf{Network Corr.}\par{}\textbf{Mean (SD)} & \textbf{Group-Average}\par{}\textbf{Network Corr.} & \textbf{Edge Corr.}\par{}\textbf{Mean (SD)} \\ 
\arrayrulecolor{black}\midrule
 \textit{(A)} Default Setting & 0.23 (0.032) & 0.331 & 0.011 (0.043) \\ 
\arrayrulecolor{black}\midrule

\arrayrulecolor{black}
\textit{(B)} Probabilistic Tractography & 0.27 (0.032) & 0.357 & 0.010 (0.057) \\ 
\arrayrulecolor{black}\midrule

\textit{(C)} FA Measurement & 0.14 (0.031) & 0.246 & 0.007 (0.039) \\ 
 \arrayrulecolor{black}\midrule

\textit{(D)} Glasser-360 Parcellation & 0.20 (0.026) & 0.325 & 0.001 (0.034) \\
\bottomrule
\end{tabular}
\caption{\label{tab:fcsc_correlation_summary} Subject network, group-average network, and edge FC-SC correlations  under different network construction settings. (A) Correlations computed using the default setting for functional and structural network construction. (B) Correlations computed using SC derived from probabilistic tractography, with all other methodological choices identical to the default setting. (C) Correlations computed using FA as the SC measure. (D) Correlations computed using the Glasser-360 brain parcellation.}
\arrayrulecolor{black}
\end{table}
 
\subsection{Edge FC-SC Correlations}
\begin{figure}[h!]
\centering
\captionsetup{font=small}
\includegraphics[width=0.7\textwidth]{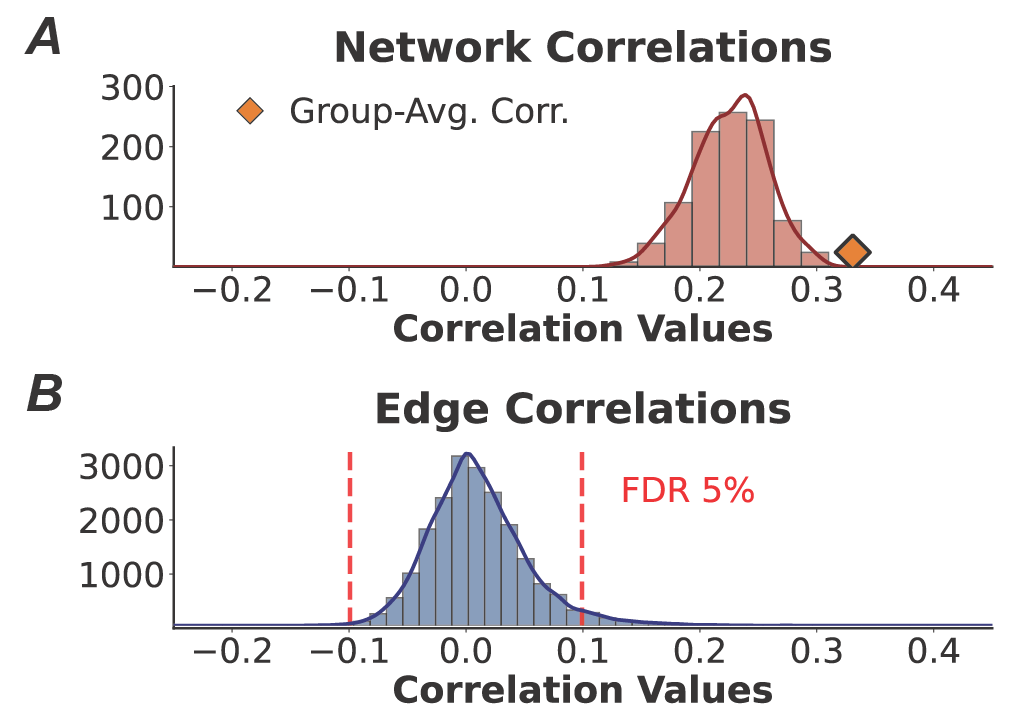}
\caption{\small
\textbf{A} The histogram of subject network FC-SC correlations calculated under the default setting. The orange diamond indicates the group-average network correlation.
\textbf{B} The histogram of edge FC-SC correlations calculated under the default setting. Red dashed lines indicate 5\% FDR thresholds.
}
\label{fig:default_setting_correlations}
\end{figure}

In contrast to moderate positive network FC-SC correlations, edge FC-SC correlations, $Corr(C_{ij}^s,\tilde{C}_{ij}^s\mid ij)$ defined in \eqref{eq:EdgeCorr}, are mostly substantially smaller. The average of edge FC-SC correlations is 0.011 with an SD of 0.043 (Table \ref{tab:fcsc_correlation_summary}.\textit{(A)}). Notably, 96.2\% of the edge correlations are statistically insignificant at the 5\% FDR threshold, as shown in Figure \ref{fig:default_setting_correlations}.\textit{B}.

\subsection{Network Construction Factors Affecting FC-SC Correlations}

Methodological choices in building functional and structural networks can affect the values of FC-SC correlations. We focused on three key factors: (1) the tractography method used to identify streamlines between brain regions, (2) the choice of quantitative measures of SC, and (3) the brain parcellation scheme used to define brain regions \citep{zhang2022quantitative}. To systematically evaluate the impact of these factors, we calculated network and edge FC-SC correlations under various methodological settings by varying one factor at a time, and compared these results to the FC-SC correlations obtained under the default setting.

\begin{figure}[ht!]
\centering
\captionsetup{font=small}
\includegraphics[width=\textwidth]{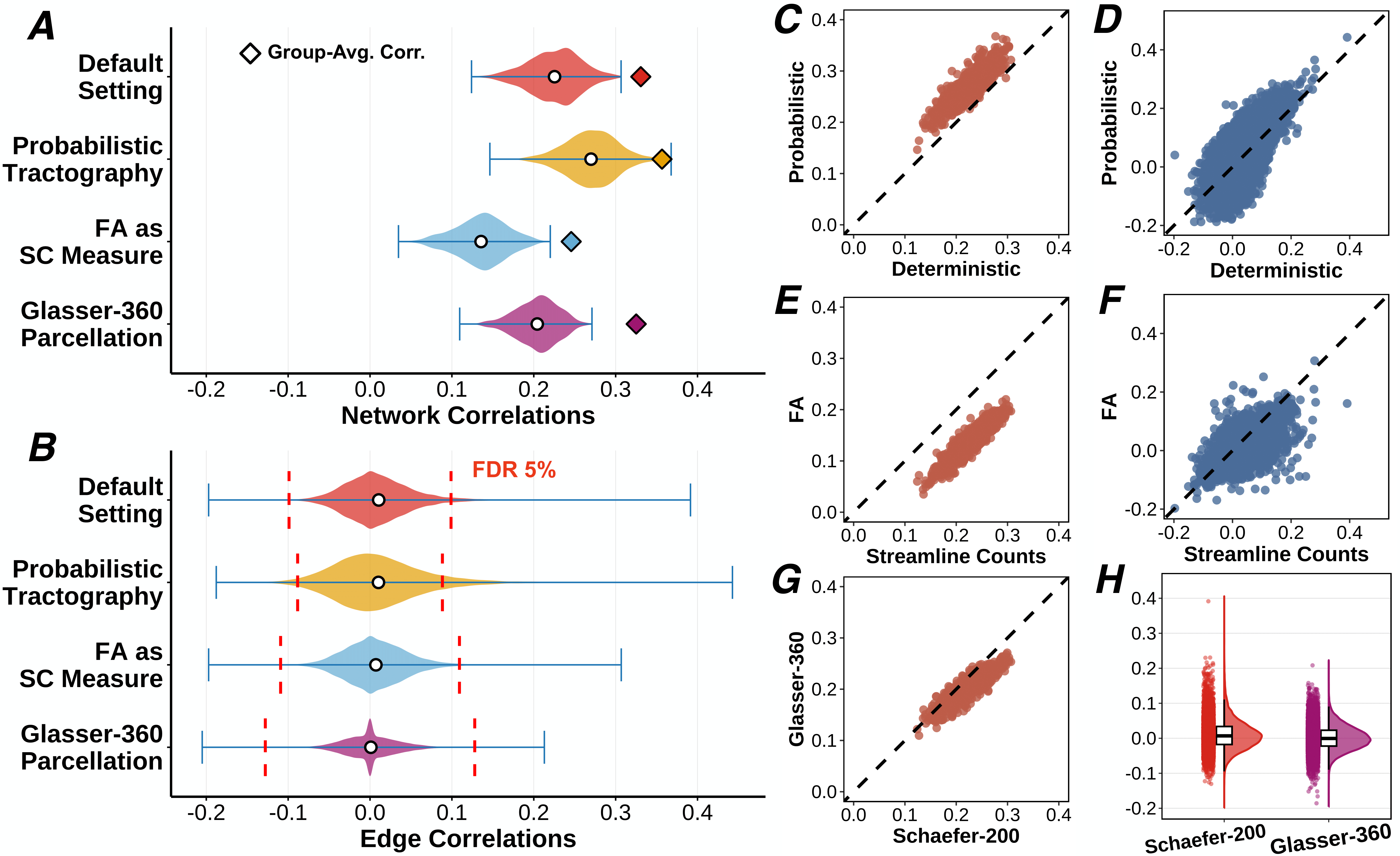}\\
\caption{\small
\textbf{A} Network FC-SC correlations across different network construction settings. White circles correspond to the means of subject network correlations, and colored diamonds indicate group-average network correlations.
\textbf{B} Edge FC-SC correlations across different network construction settings. Red dashed lines indicate 5\% FDR significance thresholds.
\textbf{C} The scatterplot comparing subject network FC-SC correlations obtained using probabilistic versus deterministic tractography.
\textbf{D} The scatterplot comparing edge correlations obtained using probabilistic versus deterministic tractography. 
\textbf{E} The scatterplot comparing subject network correlations obtained using FA-based SC versus streamline count-based SC.
\textbf{F} The scatterplot comparing edge correlations obtained using FA-based SC versus streamline count-based SC.
\textbf{G} The scatterplot comparing subject network correlations obtained using Schaefer-200 versus Glasser-360 parcellations.
\textbf{H} Violin plots comparing edge correlations obtained using Schaefer-200 versus Glasser-360 parcellations.
}
\label{fig:network_construction_choices}
\end{figure}

\subsubsection{The Tractography Method}

To assess the impact of tractography choice, we replaced our default deterministic approach with probabilistic tractography \citep{tournier2010improved}, holding all other preprocessing and network construction choices unchanged. Table \ref{tab:fcsc_correlation_summary}.\textit{(B)} summarizes the resulting FC-SC correlations.

With probabilistic tractography, the mean of subject network FC-SC correlations increases slightly to 0.27 (SD = 0.032), compared to the value under the default setting. The group‐average network correlation likewise rises to 0.357. A paired t‐test confirms that, across subjects, probabilistic-tractography‐based network correlations are statistically significantly higher than those derived from deterministic tractography (p-value $<$ 0.001), as illustrated by Figure \ref{fig:network_construction_choices}.\textit{C}. Bootstrap testing also demonstrates a significant increase in the group‐average correlation (p-value $<$ 0.001).

In contrast, edge correlations obtained using probabilistic tractography are similar to those from deterministic tractography, with a mean of 0.01 and an SD of 0.057. Still, the majority (88.8\%) of edge correlations are statistically insignificant (Figure \ref{fig:network_construction_choices}.\textit{B}). Figure \ref{fig:network_construction_choices}.\textit{D} (blue points) shows that edge correlations do not differ significantly between the two tractography methods (paired t‐test p-value $=$ 0.944).

\subsubsection{Fractional Anisotropy as SC Measure}

While streamline counts are commonly used to quantify SC between brain regions, average fractional anisotropy (FA) along these streamlines is another prevalent measure of SC strength \citep{zhang2022quantitative}.  To evaluate how the choice of SC metric influences FC-SC correlations, we repeated our analysis using FA as the SC measure while keeping all other network construction choices the same as in the default setting.

Using FA to quantify SC results in consistently smaller subject network correlations, as shown in Figure \ref{fig:network_construction_choices}.\textit{E}. The mean and SD of these correlations are 0.14 and 0.031, respectively (Table \ref{tab:fcsc_correlation_summary}.\textit{(C)}). A paired t-test also confirms that FA-based network correlations are significantly lower across subjects (p-value $<$ 0.001). Likewise, the group‐average network correlation falls to 0.246 (bootstrap p-value $<$ 0.001).

Edge FC-SC correlations are only moderately affected by the switch to FA, as shown in Figures \ref{fig:network_construction_choices}.\textit{B \& F}. The mean edge correlation remains near zero (0.007, SD = 0.039), although this small difference is statistically significant relative to streamline count-based correlations (paired t-test p-value $<$ 0.001). Moreover, 98.8\% of edges exhibit non-significant FC-SC correlations when using FA.

\subsubsection{The Brain Parcellation}
To assess the influence of the choice of the brain parcellation, we repeated all FC-SC correlation analyses using the Glasser-360 parcellation \citep{glasser2016multi} in place of the Schaefer-200 parcellation, keeping all other network construction choices unchanged. The results of the network and edge correlations are summarized in Table \ref{tab:fcsc_correlation_summary}.\textit{(D)}. Overall, the Glasser-360 parcellation produces FC-SC correlations that are slightly smaller but highly similar to those obtained using the Schaefer-200 parcellation. Specifically, the network correlations have a mean of 0.20 with an SD of 0.026, and the group-average network correlation is 0.325, which is highly similar to those under the default setting.

Figure \ref{fig:network_construction_choices}.\textit{G} shows a scatterplot comparing the network FC-SC correlations obtained using the two parcellation methods. The plot indicates that the Glasser-360 parcellation results in slightly lower network correlations for most subjects. This observation is confirmed by a paired t-test (p-value $<$ 0.001).

For edge correlations, the Glasser-360 parcellation produces statistically significantly different (two-sample t-test p-value $<$ 0.001) but practically highly similar values, as shown in Figure \ref{fig:network_construction_choices}.\textit{H}. In addition, most edges (99.9\%) have insignificant FC-SC correlations under the Glasser-360 parcellation.

\section{Discussions}

\subsection{Model-based Explanations for Divergent FC-SC Correlations}

To explain why network and edge FC-SC correlations often diverge and why they respond differently to network construction choices, we derived closed-form expressions of the two correlation types under our random effects models (Equations \eqref{eq:FC} and \eqref{eq:SC}). These expressions show that network and edge correlations depend on correlations between distinct effects in the models.

\textbf{Subject network FC-SC correlations.} For subject $s$, the network FC-SC correlation, $Corr(C_{ij}^s, \tilde{C}_{ij}^s \mid s)$, can be expressed as: 
\begin{equation}\label{eq:CorrSub}
\begin{split}
 Corr(C_{ij}^s, \tilde{C}_{ij}^s\mid s)=\rho_{\boldsymbol{\alpha}} \cdot \sqrt{P^s_{\boldsymbol{\alpha}} \cdot \tilde{P}^s_{\boldsymbol{\alpha}}}~+\varrho_{\boldsymbol{\eta}}\cdot\mbox{Sign}( \varpi^s)\cdot\mbox{Sign}( \tilde{\varpi}^s)\cdot  \sqrt{P^s_{\boldsymbol{\eta}} \cdot \tilde{P}^s_{\boldsymbol{\eta}}},\end{split}
\end{equation}
where $\rho_{\boldsymbol{\alpha}}=Corr(\alpha_{ij},\tilde{\alpha}_{ij})$, ${\varrho}_{\boldsymbol{\eta}}=Corr(\eta_{ij},\tilde{\eta}_{ij})$,
%\begin{eqnarray*}\label{eq:Ratios}
$$P^s_{\boldsymbol{\alpha}}=\frac{\tau^2_{\boldsymbol{\alpha}}}{\tau^2_{\boldsymbol{\alpha}}+(\varpi^s)^2+\sigma^2_{\boldsymbol{\epsilon}}},~~~\tilde{P}^s_{\boldsymbol{\alpha}}=\frac{\tilde{\tau}^2_{\boldsymbol{\alpha}}}{\tilde{\tau}^2_{\boldsymbol{\alpha}}+(\tilde{\varpi}^s)^2+\tilde{\sigma}_{\boldsymbol{\epsilon}}},$$
$$P^s_{\boldsymbol{\eta}}=\frac{(\varpi^s)^2\cdot \sigma^2_{\boldsymbol{\eta}}}{\tau^2_{\boldsymbol{\alpha}}+(\varpi^s)^2 \cdot \sigma^2_{\boldsymbol{\eta}}+\sigma^2_{\boldsymbol{\epsilon}}}=\frac{(\varpi^s)^2}{\tau^2_{\boldsymbol{\alpha}}+(\varpi^s)^2 +\sigma^2_{\boldsymbol{\epsilon}}},~\mbox{and}~\tilde{P}^s_{\boldsymbol{\eta}}=\frac{(\tilde{\varpi}^s)^2\cdot \tilde{\sigma}^2_{\boldsymbol{\eta}}}{\tilde{\tau}^2_{\boldsymbol{\alpha}}+(\tilde{\varpi}^s)^2 \cdot \tilde{\sigma}^2_{\boldsymbol{\eta}}+\tilde{\sigma}^2_{\boldsymbol{\epsilon}}}=\frac{(\tilde{\varpi}^s)^2}{\tilde{\tau}^2_{\boldsymbol{\alpha}}+(\tilde{\varpi}^s)^2 +\tilde{\sigma}^2_{\boldsymbol{\epsilon}}}.$$

The detailed derivation of \eqref{eq:CorrSub} based on the random effects models is provided in the supplementary file.

 Equation \eqref{eq:CorrSub} highlights two key points: First, the subject network FC-SC correlation is a weighted sum of the correlation between main edge effects ($\rho_{\boldsymbol{\alpha}}$) and the correlation between interaction edge effects in FC and SC ($\varrho_{\boldsymbol{\eta}}$), whose sign is modulated by the signs of the interaction subject effects of subject $s$ ($\mbox{Sign}(\varpi^s)$ and Sign($\tilde{\varpi}^s$)). Second,  
 the weights depend on how much of the FC and SC variances in the brain network of subject $s$ are attributable to each type of edge effects. Specifically, the ratios $P^s_{\boldsymbol{\alpha}}$ and $\tilde{P}^s_{\boldsymbol{\alpha}}$, respectively, are the proportions of FC and SC variances of subject $s$ explained by main edge effects, since the denominators $\tau^2_{\boldsymbol{\alpha}}+(\varpi^s)^2+\sigma^2_{\boldsymbol{\epsilon}}$ 
 and $\tilde{\tau}^2_{\boldsymbol{\alpha}}+(\tilde{\varpi}^s)^2 +\tilde{\sigma}^2_{\boldsymbol{\epsilon}}$ are the FC and SC variances across edges in the brain network of subject $s$, respectively.  Similarly, $P^s_{\boldsymbol{\eta}}$ and $\tilde{P}^s_{\boldsymbol{\eta}}$, respectively, are the proportions of FC and SC variances of subject $s$ explained by interaction edge effects.

We estimated all terms in Equation \eqref{eq:CorrSub} based on the fitted model parameters for each subject (estimation procedures are provided in Section \ref{sec:ParaEst}). The estimated correlations ($\hat{\rho}_{\boldsymbol{\alpha}}$ and $\hat{\varrho}_{\boldsymbol{\eta}}$) and estimated ratios ($\hat{P}^s_{\boldsymbol{\alpha}}, \breve{P}^s_{\boldsymbol{\alpha}}, \hat{P}^s_{\boldsymbol{\eta}}, \breve{P}^s_{\boldsymbol{\eta}}$) are summarized in Panel A of Table \ref{Tab:DetModelPara}.

\textbf{Group-average network FC-SC correlation.} Notably,  $\hat{\rho}_{\boldsymbol{\alpha}}$ is the group-average network FC-SC correlation and also serves as the estimate of the population-mean network FC-SC correlation.

\noindent\textbf{Edge FC-SC correlations.} For a given edge $(i,j)$ connecting regions $i$ and $j$, the edge correlation, $Corr(C_{ij}^s,\tilde{C}_{ij}^s\mid ij)$, is given by:
\begin{equation}\label{eq:CorrEdge}
    \begin{split}
  &Corr(C_{ij}^s,\tilde{C}_{ij}^s\mid ij)=\rho_{\boldsymbol{\beta}} \cdot \sqrt{P_{{\boldsymbol{\beta}},ij}\cdot \tilde{P}_{{\boldsymbol{\beta}},ij}  }+\varrho_{\boldsymbol{\varpi}} \cdot\mbox{Sign}( \eta_{ij})\cdot\mbox{Sign}( \tilde{\eta}_{ij})\cdot  \sqrt{P_{{\boldsymbol{\varpi}},ij} \cdot \tilde{P}_{{\boldsymbol{\varpi}},ij}},\end{split}
\end{equation}
where  $\rho_{\boldsymbol{\beta}}=Corr(\beta^s,\tilde{\beta}^{s})$,
$ {\varrho}_{\boldsymbol{\varpi}}=Corr(\varpi^s,\tilde{\varpi}^{s})$,
$$P_{{\boldsymbol{\beta}},ij}=\frac{\tau^2_{\boldsymbol{\beta}}}{\tau_{\boldsymbol{\beta}}^2+\eta^2_{ij}\cdot\sigma_{\boldsymbol{\varpi}}^2+\sigma^2_{\boldsymbol{\epsilon}}},~~~\tilde{P}_{{\boldsymbol{\beta}},ij}=\frac{\tilde{\tau}^2_{\boldsymbol{\beta}}}{\tilde{\tau}^2_{\boldsymbol{\beta}}+\tilde{\eta}^2_{ij}\cdot\tilde{\sigma}_{\boldsymbol{\varpi}}^2+\tilde{\sigma}^2_{\boldsymbol{\epsilon}}},$$ 
$$P_{{\boldsymbol{\varpi}},ij}=\frac{\eta_{ij}^2 \cdot\sigma_{\boldsymbol{\varpi}}^2}{\tau^2_{\boldsymbol{\beta}}+\eta^2_{ij}\cdot\sigma_{\boldsymbol{\varpi}}^2+\sigma^2_{\boldsymbol{\epsilon}}},~ \mbox{and}~
\tilde{P}_{{\boldsymbol{\varpi}},ij}=
\frac{\tilde{\eta}_{ij}^2\cdot\tilde{\sigma}_{\boldsymbol{\varpi}}^2}{\tilde{\tau}^2_{\boldsymbol{\beta}}+\tilde{\eta}^2_{ij}\cdot\tilde{\sigma}_{\boldsymbol{\varpi}}^2+\tilde{\sigma}^2_{\boldsymbol{\epsilon}}}.
$$
Equation \eqref{eq:CorrEdge} reveals that each edge correlation is a weighted sum of the correlation between main subject effects ($\rho_{\boldsymbol{\beta}}$) and the correlation between interaction subject effects in FC and SC ($\varrho_{\boldsymbol{\varpi}} $), modulated by the signs of the interaction edge effects of edge $(i,j)$ ($\mbox{Sign}(\eta_{ij})$ and Sign($\tilde{\eta}_{ij}$)). The weights for the two correlations depend on how much of the FC and SC variances across subjects at the edge are attributable to the main and interaction subject effects.

We estimated all the terms in Equation \eqref{eq:CorrEdge} using the estimated model parameters for each edge $(i,j)$. The subsequent estimated correlations ($\hat{\rho}_{\boldsymbol{\beta}}$ and $\hat{\varrho}_{\boldsymbol{\varpi}}$) and estimated ratios ($\hat{P}_{\boldsymbol{\beta},ij}, \breve{P}_{\boldsymbol{\beta},ij}, \hat{P}_{\boldsymbol{\varpi},ij}, \breve{P}_{\boldsymbol{\varpi},ij}$) are provided in Panel B of Table \ref{Tab:DetModelPara}.

\textbf{Insights from the random effects models.} Using random effects models to investigate variability in FC and SC provides key insights into the origins of the observed differences between network and edge FC-SC correlations.

First, these two types of correlations arise from  different sources of variability. Subject network FC-SC correlations depend on (main and interaction) edge effects in FC and SC, reflecting how FC and SC co-vary across edges within a subject. In contrast, edge correlations depend on (main and interaction) subject effects, capturing how FC and SC co-vary across subjects at a specific edge.

Second, the strength of FC-SC correlations depends not only on the magnitudes of correlations between corresponding effects in the FC and SC models, but also on how much these effects contribute to FC and SC variances. For example, subject network correlations are largely driven by the correlation between main edge effects ($\rho_{\boldsymbol{\alpha}}$), because the associated weight---which reflects the proportions of FC and SC variances of each subject explained by main edge effects---is high (with a 95\% confidence interval (C.I.) of [0.60,0.66]), as shown in Panel A of Table \ref{Tab:DetModelPara}. In contrast, interaction edge effects account for small fractions of FC and SC variances in each subject, resulting in a much smaller weight for the correlation between these effects ($\varrho_{\boldsymbol{\eta}}$), typically below 0.04. Consequently, their influence on subject network FC-SC correlations is minimal.

Edge FC-SC correlations, by contrast, are mostly close to zero. This is  because (main and interaction) subject effects explain only small proportions of the FC and SC variances at individual edges. As a result, the corresponding weights for the correlations between subject effects in FC and SC ($\rho_{\boldsymbol{\beta}}$ and $\varrho_{\boldsymbol{\varpi}}$) are low, typically below 10\% for most edges (Panel B, Table~\ref{Tab:DetModelPara}). This limited contribution results in weak edge FC-SC correlations.

Third, among all the correlations between corresponding effects in the FC and SC models, the correlation between main edge effects ($\rho_{\boldsymbol{\alpha}}$), equivalent to the population-mean network FC-SC correlation, is the strongest and substantially larger than the others. This finding suggests that FC-SC associations are most pronounced at the population-mean network level, whereas subject-specific FC and SC features are only weakly correlated. One possible explanation is that the estimated subject effects ($\beta^s$ and $\tilde{\beta}^s$), which are FC and SC averages across all edges for each subject, may not capture biologically meaningful variability in brain organization. Instead, these subject effects could largely reflect non-specific influences, such as motion artifacts, vascular reactivity, or features unique to acquisition protocols of fMRI and dMRI \citep{baum2018impact,zhang2022quantitative,geerligs2017challenges,champagne2022physiological}.

% Table: DetModelPara_95CI
\begin{table}[!htbp]
\centering
\small
\begin{tabular}{lcc}
\toprule
\multicolumn{3}{l}{\textbf{Panel A: Subject Network FC-SC Correlation Components}} \\
\midrule
\multicolumn{1}{l}{{\textit{\textbf{Estimates of Correlations between Effects}}}} & \multicolumn{2}{c}{\textbf{Value}} \\
\midrule
Correlation between main edge effects & \multicolumn{2}{c}{$\hat{\rho}_{\boldsymbol{\alpha}} = 0.33$} \\
Correlation between interaction edge effects & \multicolumn{2}{c}{$\hat{\varrho}_{\boldsymbol{\eta}} = -0.07$} \\
\midrule
\multicolumn{1}{l}{\textit{\textbf{Estimated Variance Proportions across Subjects}}} & \textbf{{\textbf{95\% C.I.}} for FC} & \textbf{{\textbf{95\% C.I.}} for SC} \\
\midrule
Proportion explained by main edge effects & $\sqrt{\hat{P}^s_{\boldsymbol{\alpha}}} \in [0.72, 0.78]$ & $\sqrt{\breve{P}^s_{\boldsymbol{\alpha}}} \in [0.83, 0.84]$ \\
Proportion explained by interaction edge effects & $\sqrt{\hat{P}^s_{\boldsymbol{\eta}}} \in [0.01, 0.40]$ & $\sqrt{\breve{P}^s_{\boldsymbol{\eta}}} \in [0.00, 0.14]$ \\
\midrule
\multicolumn{1}{l}{\textit{\textbf{Estimated Weights across Subjects}}} & \multicolumn{2}{c}{\textbf{{\textbf{95\% C.I.}}}} \\
\midrule
Weight for $\rho_{\boldsymbol{\alpha}}$ & \multicolumn{2}{c}{$\sqrt{\hat{P}^s_{\boldsymbol{\alpha}} \cdot \breve{P}^s_{\boldsymbol{\alpha}}} \in [0.60, 0.66]$} \\
Weight for $\varrho_{\boldsymbol{\eta}}$ & \multicolumn{2}{c}{$\sqrt{\hat{P}^s_{\boldsymbol{\eta}} \cdot \breve{P}^s_{\boldsymbol{\eta}}} \in [0.00, 0.04]$} \\
\midrule
\multicolumn{3}{l}{\textbf{Panel B: Edge FC-SC Correlation Components}} \\
\midrule
\multicolumn{1}{l}{{\textit{\textbf{Estimates of Correlations between Effects}}}} & \multicolumn{2}{c}{\textbf{Value}} \\
\midrule
Correlation between main subject effects & \multicolumn{2}{c}{$\hat{\rho}_\beta = -0.10$} \\
Correlation between interaction subject effects & \multicolumn{2}{c}{$\hat{\varrho}_\varpi = 0.12$} \\
\midrule
\multicolumn{1}{l}{\textit{\textbf{Estimated Variance Proportions across Edges}}} & \textbf{{\textbf{95\% C.I.}} for FC} & \textbf{{\textbf{95\% C.I.}} for SC} \\
\midrule
Proportion explained by main subject effects & $\sqrt{\hat{P}_{{\boldsymbol{\beta}},ij}} \in [0.55, 0.64]$ & $\sqrt{\breve{P}_{{\boldsymbol{\beta}},ij}} \in [0.05, 0.05]$ \\
Proportion explained by interaction subject effects & $\sqrt{\hat{P}_{{\boldsymbol{\varpi}},ij}} \in [0.01, 0.51]$ & $\sqrt{\breve{P}_{{\boldsymbol{\varpi}},ij}} \in [0.00, 0.34]$ \\
\midrule
\multicolumn{1}{l}{\textit{\textbf{Estimated Weights across Edges}}} & \multicolumn{2}{c}{{\textbf{95\% C.I.}}} \\
\midrule
Weight for $\rho_{\boldsymbol{\beta}}$ & \multicolumn{2}{c}{$\sqrt{\hat{P}_{{\boldsymbol{\beta}},ij} \cdot \breve{P}_{{\boldsymbol{\beta}},ij}} \in [0.03, 0.03]$} \\
Weight for $\varrho_{\boldsymbol{\varpi}}$ & \multicolumn{2}{c}{$\sqrt{\hat{P}_{{\boldsymbol{\varpi}},ij} \cdot \breve{P}_{{\boldsymbol{\varpi}},ij}} \in [0.00, 0.09]$} \\
\bottomrule
\end{tabular}
\caption{\label{Tab:DetModelPara} Panel A presents the estimates of components in the mathematical expression \eqref{eq:CorrSub} for subject network FC-SC correlations. Panel B shows the estimates of components in the mathematical expression \eqref{eq:CorrEdge} for edge FC-SC correlations.}
\end{table}

\subsection{Variance Decomposition of Total FC and SC Variability}

Our model-based analysis of network and edge FC-SC correlations reveals that the strength of these correlations strongly depends on the proportions of FC and SC variances of each subject and edge attributable to edge and subject effects, respectively.
To more comprehensively assess the relative contributions of different effects, we conducted variance decomposition of the total FC and SC variability across all subjects and edges. Detailed procedures for this analysis are provided in Section \ref{Sec:VarDec}.

As summarized in columns \textit{A} and \textit{B} of Table \ref{Tab:VarComp}, main edge effects are the dominant contributors, explaining 46.3\% of total FC variability and an even larger 70.5\% of total SC variability. Main subject effects account for a meaningful proportion of FC variability (20.2\%) but make negligible contributions to SC variability. Interaction effects explain only a small fraction of the variability in both FC and SC. Finally, residual error terms, which capture sources of variation not characterized by the random effects, account for approximately 30\% of the total variability in both FC and SC.

These results further corroborate our earlier findings: FC-SC associations are primarily driven by main edge effects, i.e., FC and SC variability across edges, while variability tied to individual subjects (subject effects) or edge–subject interactions (interaction effects) plays a much smaller role  in shaping these associations.

\begin{table}[ht!]
\centering
\setlength{\extrarowheight}{0pt}
\addtolength{\extrarowheight}{\aboverulesep}
\addtolength{\extrarowheight}{\belowrulesep}
\setlength{\aboverulesep}{0pt}
\setlength{\belowrulesep}{0pt}
\label{Tab:VarComp}
\begin{tabular}{>{\centering\hspace{0pt}}m{0.20\linewidth}|>{\centering\hspace{0pt}}m{0.09\linewidth}>{\centering\hspace{0pt}}m{0.1\linewidth}>{\centering\hspace{0pt}}m{0.15\linewidth}>{\centering\hspace{0pt}}m{0.09\linewidth}|>{\centering\hspace{0pt}}m{0.09\linewidth}>{\centering\arraybackslash\hspace{0pt}}m{0.09\linewidth}} 
\toprule
& \textit{A} & \textit{B} & \textit{C} & \textit{D} & \textit{E} & \textit{F} \\ 
\cline{2-7}

\multicolumn{1}{>{\hspace{0pt}}m{0.20\linewidth}|}{} & \textbf{FC}  & \multicolumn{3}{>{\centering\hspace{0pt}}m{0.34\linewidth}|}{\textbf{SC}} &  \textbf{FC} & \textbf{SC}  \\
 \cmidrule(lr){2-5} \cmidrule(lr){6-7}
\multicolumn{1}{>{\hspace{0pt}}m{0.20\linewidth}|}{\textbf{Variation Source}} & \multicolumn{4}{>{\centering\arraybackslash\hspace{0pt}}m{0.5\linewidth}|}{\textbf{Schaefer-200}} & \multicolumn{2}{>{\centering\arraybackslash\hspace{0pt}}m{0.2\linewidth}}{\textbf{Glasser-360}} \\ 
\cmidrule(lr){2-2}\cmidrule(lr){3-5}

\multicolumn{1}{>{\hspace{0pt}}m{0.20\linewidth}|}{} & \textbf{Default}\par{}\textbf{Setting} & \textbf{Default}\par{}\textbf{Setting} & \textbf{Probabilistic}\par{}\textbf{Tractography} & \textbf{FA} & & \\ 

 %\cmidrule(lr){3-5} 
%\cline{1-7}
%\textbf{Variation Source} &  &  &  &  &  &  \\ 
\hhline{-------}
Main edge effects\par{}$R^2(\boldsymbol{\alpha})$ & 46.3\% & 70.5\% & 88.2\% & 44.7\% & 43.6\% & 68.2\% \\
\arrayrulecolor{black}\midrule

Main subject effects\par{}$R^2(\boldsymbol{\beta})$ & 20.2\% & 0.08\% & 0.4\% & 0.3\% & 17.2\% & 0.04\% \\
\arrayrulecolor{black}\midrule

 Interaction effects\par{}$R^2(\boldsymbol{\eta \cdot \varpi})$& 3.7\% & 0.4\% & 0.4\% & 0.7\% & 4.3\% & 0.3\% \\
\arrayrulecolor{black}\midrule

Residuals\par{}$R^2(\boldsymbol{\epsilon})$ & 29.8\% & 29.0\% & 11.0\% & 54.3\% & 34.9\% & 31.4\% \\
\bottomrule
\end{tabular}
\caption{\label{Tab:VarComp} Variance decomposition of total FC and SC variability under the default and alternative network construction settings. The table summarizes the proportions of total FC and SC variability explained by main edge effects, main subject effects, interaction effects, and residuals under the default (columns A and B) and three alternative network construction settings: (1) using probabilistic tractography to identify streamlines (column C), (2) using FA as the SC measure (column D), and (3) using the Glasser-360 parcellation to determine brain regions (columns E and F).}
\end{table}

\subsubsection{Similarity of Structural Networks Across Subjects}
In addition to clarifying differences between network and edge FC-SC correlations, the variance decomposition results reveal that structural brain networks are substantially more similar across subjects than functional networks. This conclusion is supported by the observation that main edge effects account for 70.5\% of the total variability in SC, compared to 46.3\% in FC.

\begin{figure}[ht!]
\centering
\captionsetup{font=small}
\includegraphics[width=\textwidth]{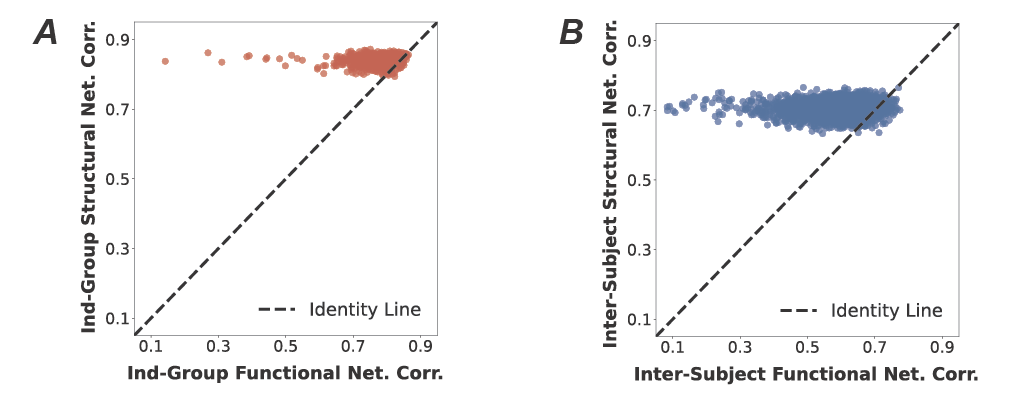}
\caption{\small Comparison of inter-subject similarity in functional and structural brain networks.
\textbf{A} The scatterplot comparing correlations between individual and group-average structural networks versus correlations between individual and group-average functional networks. Each red point represents a subject. 
\textbf{B} The scatterplot comparing correlations between structural networks of randomly selected subject pairs versus the corresponding correlations between their functional networks.  Each blue point corresponds to one pair of subjects.
}
\label{fig:population_correlation_analysis}
\end{figure}

The conclusion is further supported by Figure \ref{fig:population_correlation_analysis}.\textit{A}, which shows that correlations between each subject’s structural network and the group-average structural network typically fall between 0.8 and 0.9.  In contrast, correlations between subjects' functional networks and the group-average functional network are more dispersed, typically ranging from 0.5 to 0.8, with several falling below 0.5. Furthermore, as illustrated in Figure \ref{fig:population_correlation_analysis}.\textit{B}, correlations between structural networks of randomly selected subject pairs hover around 0.7---markedly higher than those between their corresponding functional networks.

These findings underscore the higher uniformity of structural networks compared to functional networks among healthy young adults, consistent with previous reports highlighting the relative stability of SC and more dynamic, state-dependent nature of FC \citep{Zimmermann2018Subject}. 

\subsubsection{Evaluation of Potential Family Effects}

To determine whether the high similarity observed in structural networks across subjects, particularly the large proportion of SC variability explained by main edge effects, could be attributed to the inclusion of multiple subjects from the same family, we conducted the same analysis using FC and SC data of 429 subjects from distinct families. As detailed in Supplementary Table S4, the correlations between parameter estimates and  variance decomposition results are remarkably similar to those derived from the entire dataset from the HCP. This consistency suggests that the high similarity in structural networks is not due to family-related biases in the data.

\subsection{Variance Decomposition with Alternative Network Construction}
To assess how choices in network construction influence the variability of FC and SC and their correlations, we extended our random effects modeling framework and variance decomposition analyses across alternative settings. 
This section compares results from these alternatives to the default setting, focusing on how specific choices affect the proportion of explained variability in FC and SC and, in turn, the strength of FC-SC correlations.

\subsubsection{Probabilistic Tractography}

We observed greater subject and group-average network FC-SC correlations when structural networks were constructed using probabilistic tractography rather than deterministic tractography. These increases are directly linked to a substantial increase in the proportion of total SC variability captured by main edge effects. As reported in column \textit{C} of Table \ref{Tab:VarComp}, 88\% of SC variability is attributable to main edge effects under probabilistic tractography, which is considerably higher than 71\% captured under deterministic tractography. At the same time, the proportion of unexplained SC variation (residuals) decreases sharply from 29\% to 11\%. This enhanced ability to capture SC variability across edges results in stronger network FC-SC correlations.

This improvement is likely due to fundamental methodological differences between tractography approaches. Probabilistic tractography tends to identify a broader and more distributed set of streamlines between brain regions \citep{maffei2022insights, sarwar2019mapping}, producing denser structural networks with more edges exhibiting nonzero SC values. These denser networks more effectively reflect continuous variation in SC strength across edges. In contrast, deterministic tractography often generates sparser networks with a larger number of zero-valued edges, limiting the ability to detect inter-edge variability in SC and thus constraining FC-SC associations.

Despite more SC variability captured by main edge effects under probabilistic tractography, main subject and interaction effects continue to account for only a small portion of the total SC variability. As a result, edge FC-SC correlations remain weak, similar to those under the default (deterministic) setting.

\subsubsection{FA Measurements}
The variance decomposition analysis also sheds light on why network FC-SC correlations are substantially lower when SC is quantified using FA instead of streamline counts. Specifically, the drop in network FC-SC correlations is explained by a marked reduction in the proportion of SC variability attributable to main edge effects---from 71\% under the default streamline-based SC to only 45\% with FA-based SC, as summarized in column \textit{D} of Table \ref{Tab:VarComp}. Concurrently, the proportion of unexplained SC variation increases dramatically from 29\% to 54\%, indicating that FA introduces substantially more residual variability not captured by random effects.

These findings suggest that FA-based SC measures are more susceptible to noise and confounding variability. This aligns with existing literature, which highlights that FA, as a microstructural measure, is highly sensitive to factors such as white matter geometry, crossing fibers, and partial volume effects \citep{straathof2019systematic, vos2011partial, vos2012influence, jones2013white}. Our model-based variance decomposition quantitatively confirms this limitation: much of the variability in FA-derived SC is unrelated to large-scale fiber pathways, thereby reducing its correlation with FC patterns.

As in previous settings, main subject and interaction effects explain only a minor portion of the total SC variability under FA-based SC, leading to statistically insignificant edge FC-SC correlations for nearly all edges.

%These results underscore the importance of selecting SC metrics that maximize interpretable signal-to-noise ratio and highlight streamline count as a more robust choice for assessing structure–function coupling at the network level.

\subsubsection{Glasser-360 Parcellation}

We also evaluated how brain parcellation influences FC and SC variability and their correlations by comparing the default Schaefer-200 parcellation to the higher-resolution Glasser-360 parcellation. As shown in columns \textit{E} and \textit{F} of Table~\ref{Tab:VarComp}, using the Glasser-360 parcellation results in slight decreases in the proportions of FC and SC variability explained by main edge effects, accompanied by small increases in residual variability. Consequently, both subject and group-average network FC-SC correlations are slightly lower compared to those obtained with the Schaefer-200 parcellation.

Moreover, the contributions of main subject and interaction effects remain minimal under the Glasser-360 parcellation as well. As a result, edge FC-SC correlations remain weak,  consistent with results under other settings.

Overall, the variance decomposition results suggest that the choice of brain parcellation scheme has only a small impact on the variability distribution of FC and SC and on the resulting FC-SC correlations. This robustness to the parcellation choice is consistent with prior studies investigating the influence of parcellation on function-structure coupling, network organization, and model-based prediction performance \citep{messe2020parcellation, arslan2018human, litwinczuk2024impact}.

\subsection{Methodological Advantages, Limitations, and Future Research}
The proposed random effects models offer a principled statistical framework for disentangling distinct sources of variability in FC and SC.
The models provide a flexible tool for characterizing and quantifying FC and SC variability across edges (main edge effects), variability across individuals (main subject effects), and their interactions (interaction effects).

A central methodological strength of this approach lies in its variance decomposition, which allows for systematic quantification of how each effect contributes to the total variability in FC and SC. This decomposition not only offers mechanistic insight into the function-structure relationship but also clarifies why network FC-SC correlations tend to be moderate positive, whereas edge correlations are generally weak. Specifically, the decomposition reveals how each correlation is shaped by both the strength of associations between corresponding random effects in FC and SC and the extent to which those effects account for variability in the data.

Additionally, the modeling framework supports direct comparisons across different network construction choices,including tractography methods, SC metrics, and brain parcellation schemes, by showing how these methodological decisions alter the distribution of variability across model components. This capability is particularly useful for evaluating the robustness and interpretability of FC-SC correlations under varying analytical pipelines.

Despite these strengths, several limitations remain, offering important directions for future research. First, while the models capture the patterns of weak FC-SC correlations for most edges, they fall short in explaining a small subset of edges (approximately 3.7\%) that still exhibit statistically significant FC-SC correlations. One possible explanation is that these discrepancies stem from violations of the independence and identically distributed (i.i.d.) assumptions regarding random effects. For example, main edge effects $\alpha_{ij}$ are assumed to be i.i.d. with a distribution. It is highly likely that the connectivity of some edges is unique and follows a distribution deviant from the overall trend {\citep{chumin2022cortico,jo2021diversity,zamani2022local,faskowitz2022edges}}. Future work will aim to extend the current framework by relaxing the i.i.d. assumptions and accommodating diverse distributions of different effects, and provide more accurate and interpretable modeling of complex FC-SC relationships.

Second, the present study was conducted using a cohort of healthy young adults. This homogeneity limits the generalizability of the findings. The observed similarity in structural networks across individuals may be specific to this demographically narrow population and may not generalize to other age groups or clinical populations. Future research should apply the modeling framework to more diverse samples, spanning a broader age range and clinical conditions, and examine how FC-SC variability and coupling patterns evolve across the lifespan and under pathological states.

Third, while we focused on well-established imaging processing choices to ensure reproducibility and comparability, including common tractography algorithms, SC weighting methods, and cortical parcellation schemes, other potentially influential factors were not included in the analysis. For instance, decisions such as edge pruning thresholds or the inclusion of subcortical regions may markedly alter network topology and consequently affect FC-SC correlations \citep{gu2021heritability, civier2019removal} and variance decomposition results. Future research should systematically investigate the impact of other methodological choices to better understand their influence on FC-SC relationships.

Finally, our current analysis emphasizes direct SC between region pairs and does not consider indirect structural connections---pathways through which FC may be supported. However, increasing evidence suggests that indirect SC plays a critical role in shaping functional dynamics \citep{goni2014resting, honey2009predicting}. Therefore, incorporating metrics of indirect connectivity (e.g., communicability, path transitivity, or network diffusion models) into future versions of our framework will offer a more complete evaluation of the structural underpinnings of FC. This will help refine the interpretation of FC-SC coupling and provide deeper insights into the multiscale organization of brain networks.

\section*{Data and Code Availability}

The open access HCP-YA data can be downloaded from the data management platform ConnectomeDB:
https://db.humanconnectome.org upon signing up for an account.

The code for the analyses presented in this paper is included as a compressed file in the supplementary
materials and is also available on Github (https://github.com/lypeng97/two-factor-random-effects-model).

\section*{Author Contributions}

Tingting Zhang (Research design, Methodology development, Investigation, Project administration, Visualization,
Reference review, Writing original draft), Lingyi Peng and Qiaochu Wang (Data curation and
preprocessing, Methodology implementation, Formal analysis, Investigation, Software, Validation, Visualization,
Reference review, Writing original draft), and all other co-authors (Investigation, Reference
review, Writing original draft).

\section*{Acknowledgements}
\noindent This work was supported by funding from the National Institute on Aging (NIA) grant U19AG074866.
T.Z. gratefully acknowledges the additional support by the NSF-2242568 and P01AG025204. D.L.T.
gratefully acknowledges the support by R01AG063752 and P01AG025204.

This research was supported in part by the University of Pittsburgh Center for Research Computing,
RRID:SCR 022735, through the resources provided. Specifically, this work used the H2P and HTC
clusters, which are supported by NIH award numbers OAC-2117681 and S10OD028483, respectively.

The HCP-YA data were provided, in part, by the HCP, WU-Minn Consortium (Principal Investigators:
David Van Essen and Kamil Ugurbil; 1U54MH091657) funded by the 16 NIH Institutes and Centers
that support the NIH Blueprint for Neuroscience Research; and by the McDonnell Center for Systems
Neuroscience at Washington University.

\section*{Declaration of Competing Interests}

Stacey J. Sukoff Rizzo has served as a consultant for GenPrex, Inc. and Hager Biosciences, and holds
shares in Momentum Biosciences. Gregory W. Carter has served as a consultant for Astex Pharmaceuticals.
The other authors declare no conflict of interest.

%\section*{Supplementary Material}

%Supplementary Material (created during production as a web link to online material).

%\clearpage
\printbibliography

%\appendix

%\section{Appendix}

%Appendices (optional).

%\include{SupplementaryFile}

\end{document}